%% file: main.tex
\documentclass[conference]{IEEEtran}

\usepackage{cite}
\usepackage{amsmath,amssymb,amsfonts}
\usepackage{algorithmic}
\usepackage{graphicx}
\usepackage{epstopdf}
\usepackage[dvipsnames]{xcolor}
\usepackage[final]{microtype}
\usepackage[italic]{mathastext}
\usepackage{libertine}
\usepackage[T1]{fontenc}
\usepackage{textcomp}
\usepackage[varqu,varl]{zi4}
\usepackage[all]{nowidow}
\usepackage{fancyhdr}
\usepackage{listings}
\usepackage{authblk}

\definecolor{mGreen}{rgb}{0,0.6,0}
\definecolor{mGray}{rgb}{0.5,0.5,0.5}
\definecolor{mPurple}{rgb}{0.58,0,0.82}
\definecolor{backgroundColour}{rgb}{0.95,0.95,0.92}

\lstdefinestyle{CStyle}{
    backgroundcolor=\color{backgroundColour},
    commentstyle=\color{mGreen},
    keywordstyle=\color{magenta},
    numberstyle=\tiny\color{mGray},
    stringstyle=\color{mPurple},
    basicstyle=\ttfamily\small,
    breaklines=true,
    captionpos=b,
    numbers=none,
    numbersep=1pt,
    tabsize=1,
}

\usepackage{url}
\usepackage{tikz}
\usetikzlibrary{arrows.meta,positioning,calc}

\begin{document}


\title{LLMRunner: Closed Workload Model Performance Testing Framework for LLM Servings}




\author{Arush Upadhyaya}
\author{Debadatta Mishra}
\affil{Indian Institute of Technology Kanpur}


\maketitle
\pagestyle{plain}


\begin{abstract}

Performance testing large language model (LLM) serving engines is a crucial requirement 
considering the criticality of the infrastructure for deploying conversational AI at scale.
Most of the benchmarking tools used to carry out performance evaluation of LLM serving do not
capture realistic workload orchestration unlike closed workload performance testing
infrastructure used in other domains such as web and database engines.
For example, state-of-the-art LLM benchmarks for performance testing model traffic as
streams of independent single-turn and stand-alone requests omitting the end-user
behavior such as `think time' between conversational turns.
We present LLMRunner, a performance tester for LLM serving that addresses this gap 
by providing a framework to generate realistic load consisting of multi-turn conversations 
with configurable end-user parameters such as number of users in the workload loop, 
think-time distributions etc.
Moreover, LLMRunner provides declarative matrix-sweep experiment plans, integrated GPU 
and KV-cache monitoring, and an automated artifact collection and plotting subsystem.

We use LLMRunner to comparatively analyze the performance of LLM serving
when NVIDIA CUDA Unified Virtual Memory (UVM) is used as a drop-in replacement 
memory allocator for the Key-Value cache (KV-Cache).
Our test demonstrates that UVM incurs no measurable throughput or latency penalty compared to
the baseline memory allocation technique for non-oversubscription scenarios.

\end{abstract}

\input{introduction.tex}
\input{background.tex}
\input{methodology.tex}
\input{evaluation.tex}

\input{conclusion.tex}
\bibliographystyle{IEEEtranS}
\bibliography{reference}

\end{document}

%% file: introduction.tex
\section{Introduction}
\label{sec:introduction}

Large Language Models (LLMs) have become central to a wide range of production services, 
from conversational assistants and code-generation tools to document
summarization and retrieval-augmented pipelines~\cite{llms1, llms2,llms3}.
The performance requirements of LLM inference setups
require an efficient \emph{serving} infrastructure that can process requests
in an efficient manner, mostly using specialized hardware such as general purpose GPU hardware.
As model sizes and context lengths grow, serving has become fundamentally
memory-bound.
Therefore, efficient management of GPU memory used for
data structures such as the Key-Value (KV) cache~\cite{PagedAttention}
becomes crucial to realize a low-latency and high-throughput
inference service.
Systems such as vLLM~\cite{PagedAttention} and Orca~\cite{Orca},
built on techniques like PagedAttention and continuous batching,
have shaped the current state-of-the-art setups for LLM inference.

Performance testing these LLM serving systems under conditions that faithfully
represent production workload orchestration is essential for understanding their
true performance envelope and for improving their resource management
strategies.
Most of the existing benchmark tools~\cite{MLPerf, LLMPerf, GenAIPerf} focus
on comparative study of LLM serving systems, which is useful for enhancing the
technology and its deployments.  
However, available tools fall short to model {\em realistic usage behavior} 
which is crucial to derive {\em service level agreements (SLAs)} for the end-users, 
and perform {\em capacity planning} of LLM service infrastructure.
In this paper, we address this gap through {\em LLMRunner}, a benchmarking framework
that models {\em realistic end-user behavior and workload patterns}.

To characterize the end-user performance and perform capacity planning
of the service infrastructure, closed workload performance testing models are considered
to be very useful~\cite{closedperf1, closedperf2, jmeter}.
This performance testing approach captures realistic usage patterns along with 
end-user behavior such as `think time' i.e., the time spent by the end-user to decide the 
next action.
Real-life LLM inference deployments are similar to the above mentioned closed-loop user
workload model where users send a prompt, read/process the reply, and then send
the next prompt.
Existing LLM serving benchmarks model traffic as streams of independent requests without
considering the end-user usage aspects, typically generated using stochastic models for generating 
load~\cite{MLPerf, LLMPerf, GenAIPerf}.

A crucial aspect of LLM workload is {\em multi-turn conversations} where a 
user sends a prompt, reads the reply, thinks for some interval, and then sends
a follow-up prompt again.
Evidence from large-scale conversation logs confirms the dominance 
of multi-turn conversations.
For example, WildChat~\cite{WildChat} analyzed one million real ChatGPT interaction logs 
and found that production traffic is dominated by multi-turn sessions with heavy-tailed 
prompt length distributions.
Modeling multi-turn conversations along with end-user think time
is a crucial aspect, because processing and memory resources required at the server side
are expected to be impacted due to the increased context lengths that accumulate across turns.
To the best of our knowledge, existing benchmark tools do not support
multi-turn conversations.


A realistic benchmark should consider think time and multi-turn session structure in order to 
faithfully reproduce the concurrency and memory pressure patterns that govern real serving
behavior.
{\em LLMRunner} proposes a {\em workload generator} where parameters such as
number of users in the workload loop, prompt length and think time distribution 
can be specified. 
LLMRunner also supports both single-turn open-loop traffic and full multi-turn conversations
with configurations to define the workload mix.
LLMRunner uses existing prompt dataset~\cite{WildChat} to sample and fire real prompts,
and it is possible to provide any dataset as an input to the performance benchmarking
system.
Furthermore, extensive statistics collection at the server side resources such as 
GPU/CPU memory usage, KV cache usage along with reporting are included in the proposed
benchmark suite.
We integrated LLMRunner with the vLLM~\cite{PagedAttention} serving setup,
which is capable of hosting a wide range of language models.

In LLM serving setups, there is growing interest in 
{\em flexible and efficient management} of the KV cache~\cite{PagedAttention, kvc1, kvc2}.
Many KV cache management techniques such as vLLM~\cite{PagedAttention} pre-allocate 
a fixed slab of GPU memory for the KV cache before any requests arrive, using standard 
device memory allocation APIs (e.g., {\tt cudaMalloc} in NVIDIA GPUs).
%
%
Unified Virtual Memory (UVM) for GPU devices is an attractive substitute 
for KV-cache management because it removes the hard coupling between the allocation mechanism 
and data placement at different times during the workload execution.
UVM enables exploring future strategies such as demand-driven migration 
and capacity over-subscription in a transparent manner.
As a case-study of LLMRunner, we study the feasibility of using UVM-based KV cache to be
a drop-in replacement of existing pre-allocation strategies.
This is an interesting exploration because analyzing the impact on the user response time in presence
of different mix of realistic conversations can lead to further analysis and optimizations
for UVM-backed LLM systems.
We added support for KV cache allocation using NVIDIA UVM APIs into vLLM~\cite{PagedAttention} 
KV cache management module.
Performance analysis with LLMRunner for augmented vLLM~\cite{PagedAttention} serving infrastructure
shows that there is no noticeable impact on steady-state performance when using UVM memory
allocation strategy for KV Cache.

In this paper, we make the following contributions:

\begin{itemize}
    \item LLMRunner, a performance benchmarking tool for LLM systems
	  modelling closed-loop end-user behavior with realistic workload generation
	  capabilities.
    \item Configuration support to define flexible execution plans, resource usage 
	  monitoring, and an automated artifact collection and reporting
	  system.
    \item Study the performance of UVM-backed KV cache management demonstrating
	  the utility of LLMRunner.
\end{itemize}	    

%
%

%% file: background.tex
\section{Background and Related Work}
\label{sec:background}

\subsection{LLM Serving Systems}


Large Language Models are neural networks trained on large text corpora
to predict and generate natural language.
The transformer architecture~\cite{LLMServingSurvey}, with its multi-head
self-attention mechanism, underlies virtually all modern LLMs.
During inference, attention over the full preceding context is required for
every generated token; this computation is \emph{autoregressive}---the
model produces one token at a time, and each token depends on all tokens
generated previously.

\paragraph{Continuous batching}
Early LLM serving waited for an entire batch to complete before admitting new
requests.
Orca~\cite{Orca} introduced \emph{iteration-level scheduling} (continuous
batching)---the scheduler can insert a new request into the batch at any
decoding step, replacing a completed sequence immediately.
This approach improves GPU resource utilization.

\paragraph{The KV cache}
To avoid recomputing attention at every decoding step, serving systems cache
the key and value projections for each prior token in GPU memory.
These \emph{Key-Value (KV) tensors} accumulate throughout a generation.
For a response of $T$ tokens on a model with $L$ layers and hidden dimension $d$,
the KV cache for a single request occupies $O(T \times L \times d)$ memory.
With increasing number of requests, this aggregate can reach tens of
gigabytes, making the KV cache the dominant consumer of GPU memory.

\paragraph{PagedAttention}
Prior systems pre-allocated a contiguous slab of GPU memory per request at
admission time, causing heavy internal fragmentation.
vLLM~\cite{PagedAttention} addressed this with \emph{PagedAttention} where the KV
state is stored in fixed-size non-contiguous \emph{blocks}, allocated on
demand from a shared pool, analogous to virtual-memory paging in operating
systems.
The benefits of this technique are reduction of memory fragmentation
and sharing of KV cache blocks across requests.

\subsection{Performance Benchmarking of LLM Serving}

\paragraph{Open-loop benchmarking}
In an \emph{open-loop} model, the client issues requests at a configured
rate without waiting for prior responses; load is controlled by adjusting the
rate of request generation.
Typically, rate of requests is derived from a probability distribution.
This is the predominant model used by existing LLM benchmarks~\cite{MLPerf,
LLMPerf, GenAIPerf}.

\paragraph{Closed-loop benchmarking and think time}
In a \emph{closed-loop} performance test model, a fixed number of 
simulated users issue one request at a time and wait for the response 
before sending the next.
Between requests, every user pauses for a \emph{think-time} interval---the time
spent reading the reply and composing a follow-up.
The think time may not be fixed and can be modelled from a distribution.
Typical closed-loop performance testing tools for web-based systems
include ramp-up and steady-state phases~\cite{jmeter}.
The ramp-up phase is used to control the rate of addition of new users 
into the system. 
In the steady state, all users in the system execute the closed loop
as explained above.
The primary benefits of closed-loop performance testing is the focus
on modelling the end-user behavior accurately to provide insights 
into system performance and capacity planning~\cite{little}.
%
%


\subsection{CUDA Unified Virtual Memory}

CUDA Unified Virtual Memory (UVM)~\cite{uvm1, uvm2} is a memory-management feature of the NVIDIA
CUDA runtime~\cite{uvmnvidia} that exposes a single virtual address space accessible from both
the CPU and the GPU.
With NVIDIA UVM, allocation made with \texttt{cudaMallocManaged} can be accessed
using the same virtual address from the CPU and GPU.
The GPU hardware and driver subsystems transparently
migrate physical pages between host DRAM and GPU device memory on access.
In contrast, the classic CUDA allocator (\texttt{cudaMalloc}) pins pages in device memory
at allocation time.

From the programmer perspective, the data used for GPU computation (i.e., accessed from GPU kernels)
need to be copied to the device memory allocated using classic APIs.
With UVM, that data can be populated directly from the CPU and can be 
accessed from GPU kernels without requiring programmer intervention. 
UVM replaces this with a demand-paging model---when the GPU accesses a page
residing in host memory, a page-fault triggers its migration to device memory;
pages can be evicted back to the host under memory pressure.
Within normal GPU memory utilization bounds, where the working set fits within
physical device memory, UVM-managed allocations behave similar to standard device allocations, 
except for the overheads introduced during the first access from the GPU.
The classical GPU memory allocation APIs can not support memory over-subscription
whereas UVM can support due to the added virtualization layer.

\subsection{Related Work}

End-to-end performance benchmarking systems for different applications such as web-based
applications~\cite{closedperf1, closedperf2, jmeter, lrp}, IoT application~\cite{openrist}, 
cloud setups~\cite{faasperf, cloudperf1, cloudperf2} have been proposed and used for both 
in furthering research in the respective domains and enterprise deployments.
To the best of our knowledge, LLMRunner is the first end-to-end closed workload performance
testing tool for LLM setups.
We draw inspiration from performance testing tools such as Apache Jmeter~\cite{jmeter}
to provide a open source performance tool for LLM workloads.

Performance testing for LLM servings with simple rate-based client workload generators such as 
LLMPerf~\cite{LLMPerf}, GenAIPerf~\cite{GenAIPerf}, and similar custom performance tests are used in many
recent research proposals~\cite{PagedAttention, kvco1, kvco2}.
These techniques do not support realistic end-user modelling and comprehensive test scenario generation 
with flexible configurations.
The MLPerf inference benhmark~\cite{MLPerf} proposes a comprehensive benchmarking ecosystem 
for general machine learning workloads.
LLMRunner complements this framework with closed workload model performance benchmark support
with realistic workloads.

%% file: methodology.tex
\section{LLMRunner}
\label{sec:methodology}

\subsection{Design Goals}

LLMRunner is designed to meet the following three requirements.

\textit{Workload realism.} The framework supports full multi-turn conversation
pools with configurable per-user think-time distributions, enabling faithful
modeling of end-user interaction and the KV-cache accumulation patterns it
induces.

\textit{Systematic sweep.} An entire experiment is expressed as a single
declarative YAML \emph{plan} that the harness expands into a Cartesian
product of runs, launches each in isolation, and collects all artifacts
without manual intervention.

\textit{Reproducibility.} Every run receives a stable, content-derived
identifier so that runs from different sessions that share the same
configuration can be matched for comparison without directory-encoded
parameters.

\subsection{System Architecture}

\begin{figure*}[!t]
\centering
\begin{tikzpicture}[
  stage/.style={
    draw=black!55, rectangle, rounded corners=4pt,
    minimum height=2.1cm, text width=2.8cm,
    align=center, font=\small
  },
  arr/.style={-{Stealth[length=6pt,width=4.5pt]}, thick, draw=black!55},
  lbl/.style={font=\scriptsize\itshape, midway, above=3pt},
]
  \node[stage, fill=blue!10] (plan) at (0,0) {
    \textbf{YAML Plan}\\[4pt]
    \scriptsize - Models \\ - Workload \\ - Sweep Matrix \\ - Monitoring Plan
  };
  \node[stage, fill=yellow!25, right=1.4cm of plan] (expand) {
    \textbf{Expand}\\[4pt]
    \scriptsize - Validate Plan \\ - Generate RunSpecs with unique IDs
  };
  \node[stage, fill=green!12, right=1.4cm of expand,
        minimum width=3.5cm, text width=3.3cm] (exec) {
    \textbf{Execute}\\[4pt]
    \scriptsize - Launch Serving Instance \\ - Generate Load from Spec \\ - Monitor Resource Usage
  };
  \node[stage, fill=gray!15, right=1.4cm of exec] (output) {
    \textbf{Artifacts}\\[4pt]
    \scriptsize -  Result Collection \\ - Analysis  \\ - Reporting
  };
  \draw[arr] (plan.east)   -- node[lbl] {Validate}  (expand.west);
  \draw[arr] (expand.east) -- node[lbl] {RunSpecs}   (exec.west);
  \draw[arr] (exec.east)   -- node[lbl] {Output}    (output.west);
\end{tikzpicture}
\caption{LLMRunner four-layer execution pipeline. A declarative YAML plan is
validated and expanded into a Cartesian-product matrix of run specifications,
each assigned a stable 12-character SHA-1 identifier.
Runs execute serially---each launches a fresh serving instance
alongside a load generator and concurrent resource monitor; all artifacts
are written to an ID-keyed output directory.}
\label{fig:pipeline}
\end{figure*}
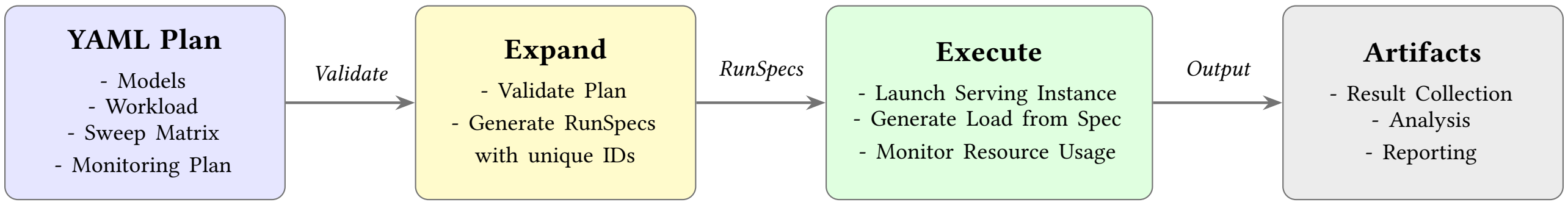

LLMRunner is organized into four layers that execute sequentially for each
run in the expanded matrix. Figure~\ref{fig:pipeline} summarizes the
end-to-end pipeline.

\textbf{Plan layer.}
A single YAML plan file defines the scope for an experiment.
It declares server configuration (model, GPU memory fraction,
allocator type), workload parameters (prompt classes, datasets,
think-time distributions), a \texttt{matrix} section listing all
values to sweep, monitoring settings, and an output directory template.

\textbf{Expansion layer.}
At parse time the harness computes the Cartesian product of all matrix
dimensions and instantiates one frozen run specification per combination.
Each specification bundles server configuration, workload sizing, load
profile, and optional multi-turn parameters, fully describing one isolated
run.
Each specification receives a 12-character SHA-1 identifier derived from
its content, making run IDs stable across re-runs and sessions.

\textbf{Execution layer.}
For each run specification, the harness {\em (i)} starts \texttt{vllm serve}
with the correct environment variables including \texttt{VLLM\_USE\_UVM},
{\em(ii)} polls the server's health endpoints until it is ready,
{\em(iii)} launches the load generator, {\em(iv)} runs a background
monitor that concurrently samples GPU, CPU, and KV-cache metrics, and
{\em(v)} shuts the server down on completion.
Runs are fully serialized: only one server and one load generator are
active at any time, ensuring unambiguous resource measurements.

\textbf{Artifact layer.}
Every run directory is self-contained, containing a complete configuration
record, aggregate statistics, per-request latency samples, system resource
time series, and KV-cache occupancy time series.
Failed runs still retain whatever partial data was collected before failure,
enabling post-mortem analysis.

\subsection{Plan YAML Specification}
\label{sec:plan-yaml}

A complete experiment is expressed in a single declarative YAML plan.
Listing~\ref{lst:full-plan} shows a representative plan covering all
major configuration sections; the discussion below walks through each
section.

\begin{lstlisting}[style=CStyle, numbers=left, caption={Representative LLMRunner plan demonstrating all major configuration sections.}, label={lst:full-plan}]
meta:
  name: llmrunner-example

globals:
  host: localhost
  port: 8000
  output_dir: output/

server:
  serve_args:
    dtype: auto
    max_model_len: 32768

monitoring:
  interval_s: 5
  gpu:
    headers: [gpu_util, gpu_mem_used]
  cpu:
    headers: [cpu_us, cpu_id, mem_free]
  kv_cache:
    headers: [kv_cache_usage_perc, num_requests_running]

workload:
  mode: multi_turn
  dataset:
    type: sonnet
  prompt_classes:
    Qwen__Qwen3-0.6B:
      small:
        input:
          dist: {type: normal, params: {mu: 512, sigma: 128}}
        output:
          dist: {type: normal, params: {mu: 128, sigma: 32}}
        num_prompts: 5000
        think_time:
          type: distribution
          dist: {type: lognormal, params: {mu: 2.0, sigma: 0.5}}
        num_conversations: 50
        max_turns: 8

matrix:
  models: [Qwen/Qwen3-0.6B]
  prompt_classes: [small]
  utils: [0.9]
  uvm: [base, uvm]
  load:
    mode: threads
    threads: [32, 48, 64, 80]
	# spacing: 5 # only for rate mode
  serve_sweep:
    mode: product
    args:
      max_num_batched_tokens: [8192, 16384]
      max_num_seqs: [256, 512]
  bench_variants:
    mode: product
    args:
      duration_s: [60, 120]
\end{lstlisting}

\textit{Infrastructure} (lines~1--21).
The \texttt{globals} block defines the endpoint for the
vllm server.
The \texttt{server} block controls the \texttt{vllm serve} life-cycle and
accepts arbitrary server flags under \texttt{serve\_args}, allowing users
to specify arguments such as data type or context length to the vllm server.

The \texttt{monitoring} block (lines~14--21) sets the sampling interval
and selects which metrics to record; separate sub-blocks configure GPU,
CPU, and KV-cache collection, each accepting \texttt{all} to capture
every available signal or an explicit header list for targeted collection.

The \texttt{workload} block (lines~23--39) declares
the conversation mode, dataset backend, and prompt-class definitions.
LLMRunner supports two conversation modes---{\em (i)} Single-turn
mode ({\tt single\_turn}) where stand-alone prompts are generated,
and {\em (ii)} Multi-turn mode ({\tt multi\_turn}) where follow-up
prompts are generated.
In \texttt{single\_turn} mode the load generator
draws prompts from the dataset and issues them as independent one-shot
completions.
In \texttt{multi\_turn} mode (line~24) the harness maintains
a pool of active conversations, where each worker appends the next user
turn to the accumulated history, sends the full context to the server,
receives the assistant reply, and appends it before issuing the next
turn---faithfully modeling the KV-cache accumulation that arises in
extended dialogue.
The dataset backend (lines~25--26) supplies the raw prompt text and seed
conversations.
The backend dataset can be---{\em (i)} synthetic token sequences for
controlled experiments, {\em (ii)} real-world conversation corpora, 
{\em (iii)} local files with user-configurable field mappings.
The choice of backend influences both the vocabulary and the turn structure 
of the generated load.

\textit{Prompt classes} (lines~27--39) are model-specific: the outer key
uses the model name with \texttt{/} replaced by \texttt{\_\_}, ensuring
each class carries token-length expectations appropriate for that
model.
A plan may define classes for multiple models.
Within each class, token lengths may be specified as statistical
distributions (e.g., normal or lognormal, as on lines~30--33) or derived
from the dataset by sampling real prompts within a configured length
range.
The dataset backend and the prompt class together determine both
the vocabulary and the length profile of each request.
For multi-turn workloads, \texttt{num\_conversations} and
\texttt{max\_turns} (lines~38--39) bound the pool size and conversation
depth.
The \texttt{think\_time} parameter (lines~35--37), which can be a constant, a sampled
distribution, or timestamps replayed from a dataset, controls the
inter-turn delay per worker.
Short think times sustain rapid KV-cache re-access; longer values allow
eviction before the next turn, reproducing realistic memory-pressure
patterns under different usage scenarios.

\textit{Sweep matrix} (lines~41--50). The \texttt{matrix} block drives the experiment sweep.
Different dimensions of the matrix---model identifiers, prompt class names, GPU memory
utilization fractions, and allocator type---are expanded into a Cartesian
product.
The total number of experiment runs is the product of cardinalities of
all dimensions.
The \texttt{load} sub-block selects the load generation mode---{\em (i)} Rate mode, and
{\em (ii)} Threads mode.
In threads mode (line~48 in the example), the user-count sweep is
specified directly via the \texttt{threads} list (line~49); the harness
runs one closed-loop test per concurrency level.
Rate mode is configured with a per-(model, class) maximum rate value
declared in a \texttt{max\_rate\_map} sub-block and an integer
\texttt{spacing} parameter (line~50) that asks the harness to linearly
space \texttt{spacing} rates from $r_{\max}/4$ to $r_{\max}$ for each
(model, class) pair.
An explicit rates list can also be provided to override the generated sweep.
The closed-loop semantics of threads mode are detailed in
Section~\ref{sec:load-modes}.

\subsection{Orchestrating Multi-Configuration Sweeps}
\label{sec:variants}

Two orthogonal variant mechanisms extend the core sweep matrix without
restructuring the plan.

\textit{Server variants} (lines~51--55) restart the LLM serving with
different flags for each variant, enabling the same workload to be driven
against multiple server configurations in a single plan.
The \texttt{serve\_sweep} block auto-generates variants as a Cartesian
product (\texttt{mode: product}) or pairwise zip (\texttt{mode: zip})
over the specified argument lists.
In Listing~\ref{lst:full-plan}, \texttt{mode: product} produces all four
combinations of batch token limit and maximum sequence count;
\texttt{mode: zip} would instead pair elements positionally.
For experiments requiring hand-crafted configurations, named variants may
be declared explicitly under \texttt{serve\_variants} without a sweep
structure.

\textit{Bench variants} (lines~56--59) override client-side load
parameters---such as run duration or max active requests---and run the
specs with various permutations of the specified variants, supporting
\texttt{zip} and \texttt{product} modes analogously to server variants.
This allows modelling, for example, different numbers of maximum active
users when using rate mode.

\subsection{Load Generation}
\label{sec:load-modes}

\begin{figure}[t]
\centering
\begin{minipage}[b]{0.48\columnwidth}
\centering
\begin{tikzpicture}[
  arr/.style={-{Stealth[length=3.5pt,width=2.5pt]}, thin},
  scale=0.95
]
  \draw[arr] (-0.1,0) -- (3.75,0);
  \node[right, font=\scriptsize] at (3.80,0) {$t$};
  \foreach \x in {0.25, 0.68, 1.65, 2.05, 3.10} {
    \fill[blue!60] (\x,0) circle (1.5pt);
    \draw[arr, blue!70] (\x,0) -- (\x,0.75);
  }
  \draw[<->, >=Stealth, very thin] (0.25,-0.28) -- (0.68,-0.28)
      node[midway, below, font=\scriptsize] {$t_1$};
  \draw[<->, >=Stealth, very thin] (0.68,-0.28) -- (1.65,-0.28)
      node[midway, below, font=\scriptsize] {$t_2$};
  \node[font=\scriptsize, align=center] at (1.90,-0.88)
      {$t_i \!\sim\! \operatorname{Exp}(\lambda),\;\mathbb{E}[t_i]\!=\!\tfrac{1}{\lambda}$};
  \node[draw, fill=gray!15, rounded corners=2pt, font=\scriptsize,
        minimum width=3.0cm, minimum height=0.35cm]
      at (1.70,1.10) {vLLM Server};
\end{tikzpicture}\\[3pt]
\footnotesize (a)~Rate mode (open-loop)
\end{minipage}
\hfill
\begin{minipage}[b]{0.50\columnwidth}
\centering
\begin{tikzpicture}[
  tbox/.style={draw, rounded corners=2pt, fill=blue!15,
               minimum width=0.72cm, minimum height=0.30cm,
               font=\tiny, inner sep=2pt},
  arr/.style={-{Stealth[length=3pt,width=2pt]}, thin},
]
  \node[tbox][font=\scriptsize] (iss) at (0.50,0) {Issue};
  \node[tbox][font=\scriptsize] (wt)  at (1.50,0) {Wait};
  \node[tbox][font=\scriptsize] (th)  at (2.50,0) {Think};
  \draw[arr] (iss.east) -- (wt.west);
  \draw[arr] (wt.east)  -- (th.west);
  \draw[arr] (th.south) .. controls +(0,-0.42) and +(0,-0.42) .. (iss.south);
  \node[font=\scriptsize] at (1.50,0.42) {$\times N$ workers};
  \draw[gray!55, dashed, rounded corners=3pt]
      (-0.10,-0.55) rectangle (3.10,0.30);
\end{tikzpicture}\\[3pt]
\footnotesize (b)~Threads mode (closed-loop)
\end{minipage}
\caption{Load generation modes. (a)~Rate mode models a Poisson arrival
process: inter-arrival times $t_i \sim \operatorname{Exp}(\lambda)$ are
i.i.d.\ with mean $1/\lambda$, so consecutive gaps $t_1, t_2, \ldots$
vary randomly at the target rate $\lambda$.
(b)~Threads mode runs $N$ concurrent workers, each cycling
Issue\,$\to$\,Wait\,$\to$\,Think before re-submitting, bounding effective
concurrency to $N$ and self-regulating the arrival rate.}
\label{fig:load-modes}
\end{figure}
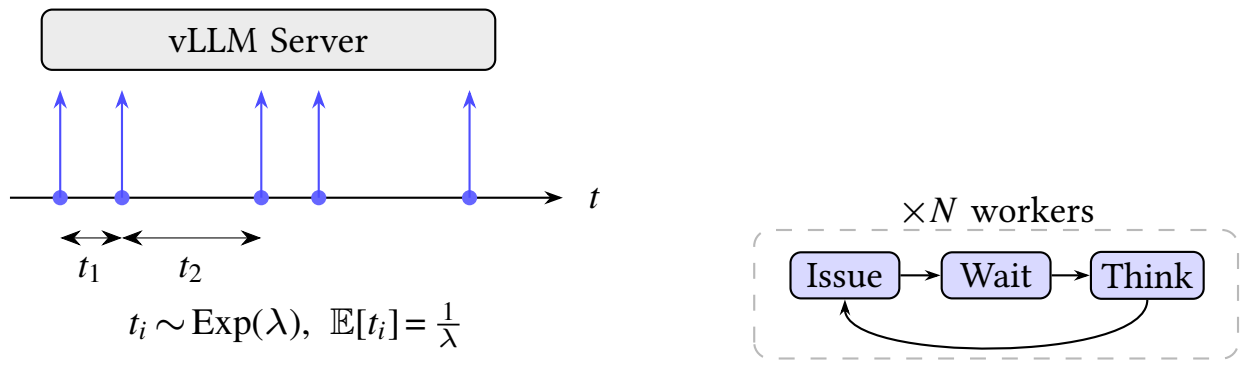

LLMRunner provides two complementary load generation modes, illustrated in
Figure~\ref{fig:load-modes}.

\subsubsection{Rate Mode (Open-Loop)}

In rate mode the client issues requests as a Poisson process at a
target rate $\lambda$ requests per second, with a configurable
maximum in-flight cap.
Inter-arrival times are i.i.d.\ draws from an exponential distribution:
\[
    t_i \;\sim\; \operatorname{Exp}(\lambda), \qquad \mathbb{E}[t_i] = \frac{1}{\lambda}
\]
Because the client does not wait for prior requests to complete, rate mode
is open-loop: queue depth grows with load until saturation.
This is the standard choice for throughput-latency sweeps and for
identifying the saturating rate under a given configuration.

\subsubsection{Threads Mode (Closed-Loop)}

In threads mode, $N$ concurrent worker threads each execute the loop:
issue a request, wait for the reply, sleep for a sampled think time, and
repeat.
The effective arrival rate is:
\[
    \lambda_{\text{eff}} = \frac{N}{\bar{T}_{\text{resp}} + \bar{T}_{\text{think}}}
\]
where $\bar{T}_{\text{resp}}$ is mean response time and
$\bar{T}_{\text{think}}$ is mean think time.
Threads mode is closed-loop: effective concurrency is bounded by $N$,
and queue depth cannot grow without bound.
This faithfully models interactive users whose send rate self-limits when
the server is slow, and sweeping thread counts yields a throughput-scaling
curve at naturally regulated concurrency.
Think time may be a fixed constant, a sampled distribution (as configured
in the prompt class), or replayed from per-request timestamps in a dataset.

\subsection{Output Structure and Run Identifiers}
\label{sec:output}

All artifacts are organized under a timestamped session directory.
Each run is stored in a subdirectory named by its 12-character SHA-1
identifier, derived from the full run configuration: model, prompt class,
utilization target, allocator type, load profile, and server and bench
arguments.
The directory contains a complete configuration record, aggregate
statistics, per-request latency samples, system resource time series,
and KV-cache occupancy time series.
Re-running the same plan produces identical run IDs, allowing artifacts
from two sessions to be compared by matching identifiers without any
external bookkeeping.

\subsection{Plotting Subsystem}
\label{sec:plotting}

The plotting subsystem converts raw artifact directories into
presentation-ready figures via \texttt{gnuplot} without manual data
extraction.
Two commands are provided: one plots aggregate result metrics (such as
throughput and latency percentiles) against a sweep axis with a second
configuration field encoded as line color; the other plots monitor
time series (GPU utilization, KV-cache occupancy) against elapsed time.
Both commands support comparing runs within a single session by a
configuration field, or overlaying the same run ID across two or more
session directories for direct baseline-versus-modified comparison.
All axis labels, titles, and legend entries are made human readable
by default making the figure ready for reporting without any post-processing.


%

%% file: evaluation.tex
\section{Evaluation}
\label{sec:evaluation}

\subsection{Experiment Setup} 
\label{sec:setup}
We ran all experiments on a Ubuntu Linux server 
with 32-core Intel(R) Xeon(R) Silver 4208 CPU, 128GB memory. 
We performed experiments using NVIDIA A40 GPU with 48GB memory and driver version 580.65.06 
For all experiments, vLLM serving setup was used as the LLM backend engine with 
Qwen/Qwen3-0.6B model of size 1.12GiB.
We modified the vLLM serving setup to add support for UVM allocation and exposed
the allocation method (Base and UVM) as a new parameter for the vLLM service. 
{\em Base} mode uses \texttt{cudaMalloc} (the default vLLM allocator) whereas 
{\em UVM} mode uses \texttt{cudaMallocManaged}.

While demonstrating the supported feature set of LLMRunner, we also 
evaluate whether replacing the standard KV-cache allocator
in vLLM with CUDA Unified Virtual Memory incurs any measurable throughput
or latency penalty under different complementary load modes.
GPU memory utilization is fixed at $u = 0.9$ for all runs, so the model and KV-cache
pool occupies 90\% of the total device memory size and the working set fits entirely within
physical GPU DRAM.
A single LLMRunner plan sweeps both allocator types (\S\ref{sec:plan-yaml}),
ensuring that each run pair is launched under identical conditions.

\subsection{Sweeping the Rate Mode}
\label{sec:eval-rate}

In rate mode (open-loop), LLMRunner issues requests as a Poisson process
at a target rate $\lambda$, sweeping five points from $\lambda_{\max}/4$
to $\lambda_{\max}$ for each prompt class.
Table~\ref{tab:config-rate} lists the token-length distributions and
offered-rate sweep used for this experiment.

\begin{table}[t]
\caption{Rate mode configuration. Token lengths are drawn from normal
  distributions parameterized as $\mathcal{N}(\mu, \sigma^2)$; five
  offered rates are linearly spaced from $\lambda_{\max}/4$ to
  $\lambda_{\max}$ per prompt class.}
\label{tab:config-rate}
\centering
\setlength{\tabcolsep}{4pt}
\begin{tabular}{|l|c|c|}
\hline
 & \textbf{Small} & \textbf{Medium} \\
\hline
Input tokens  & $\mathcal{N}(64,\,8^2)$   & $\mathcal{N}(128,\,32^2)$ \\
\hline
Output tokens & $\mathcal{N}(64,\,8^2)$   & $\mathcal{N}(128,\,32^2)$ \\
\hline
$\lambda_{\max}$ (req/s) & 120 & 84 \\
\hline
Offered rates (req/s) & 30, 53, 75, 98, 120 & 21, 37, 53, 68, 84 \\
\hline
\end{tabular}
\end{table}

\begin{figure}[tp]
\centering
\includegraphics[width=0.81\linewidth]{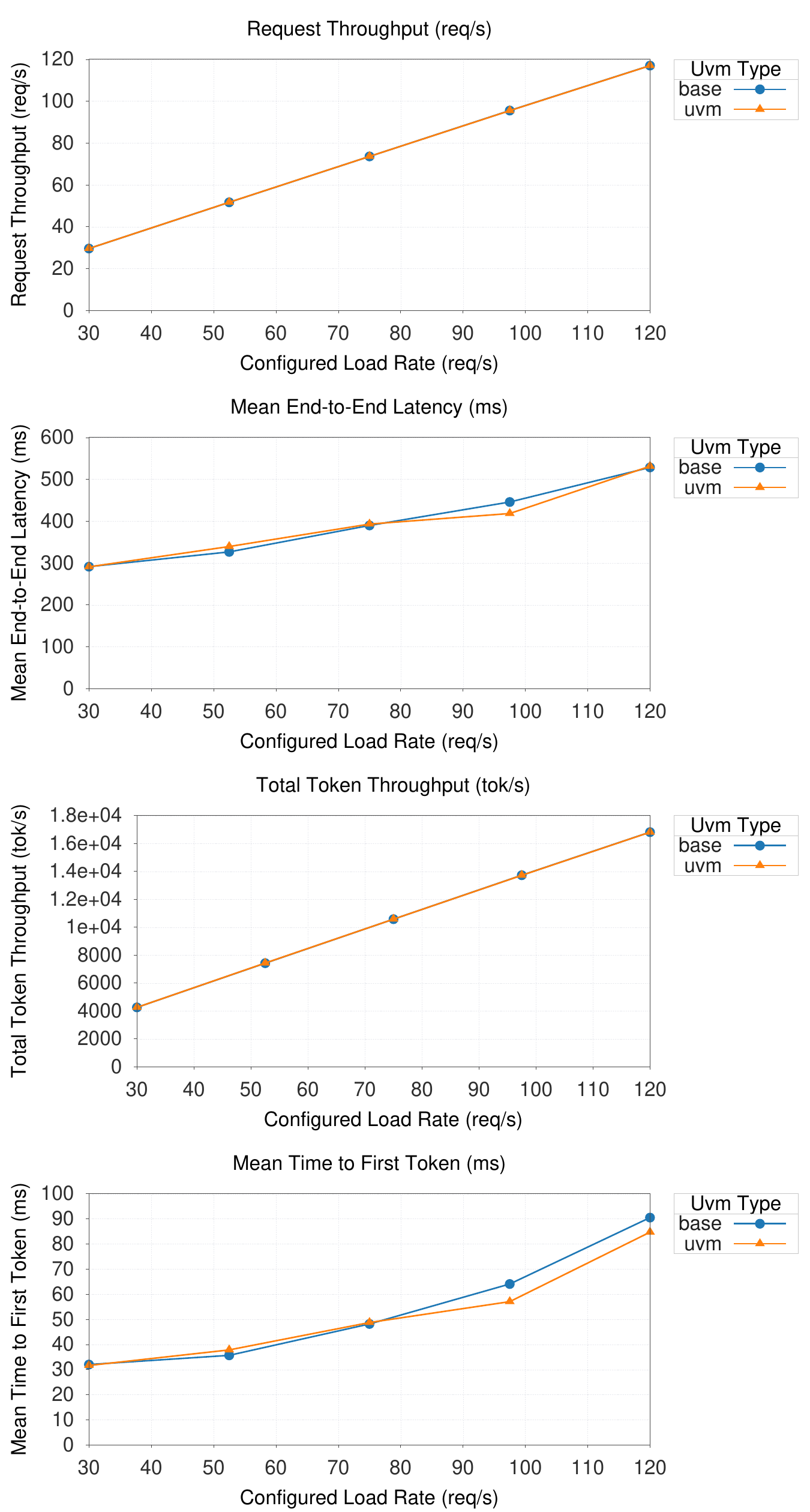}
\caption{Rate mode: small prompt class ($\sim$64\,in / $\sim$64\,out tokens,
  $u{=}0.9$). Mean end-to-end latency (left axis) and achieved throughput
  (right axis) versus offered rate. Base and UVM traces overlap throughout.}
\label{fig:rate-small}
\end{figure}

Figures~\ref{fig:rate-small} shows the performance---request throughput, mean end-to-end latency,
token generation throughput and mean time to generate first output token.
Performance of Base and UVM modes result in similar performance
in terms of all reported metric.
The average end-to-end latency increased from $\sim$300ms to $\sim$500ms
when the rate increased from 30 requests/sec to 120 requests/sec.
Similarly, mean time to generate first token was observed to be $\sim$30ms to $\sim$90ms
across different rates produced during the sweep.

\begin{figure}[t]
\centering
\includegraphics[width=0.81\linewidth]{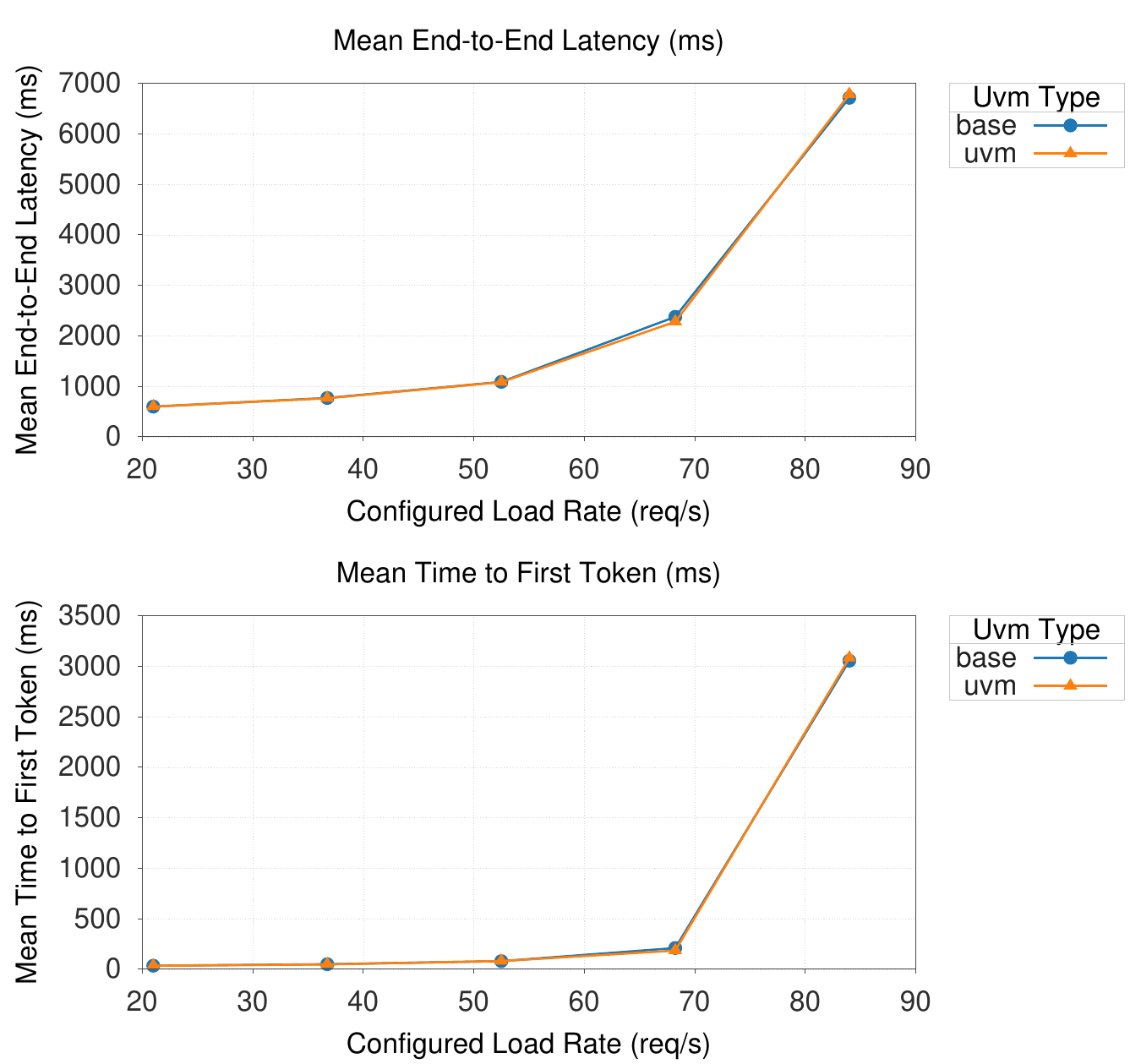}
\caption{Rate mode: medium prompt class ($\sim$128\,in / $\sim$128\,out tokens,
  $u{=}0.9$). Base and UVM traces remain coincident up to near-saturation
  rates.}
\label{fig:rate-medium}
\end{figure}

With medium class (Table~\ref{tab:config-rate}), we can observe a saturation of 
the LLM serving system when the request rate is more than 53 req/sec, both with Base and UVM (Figure~\ref{fig:rate-medium}).
As shown in the figure, `knee point' of the system results in sharp (exponential)
increase in end-to-end response time from 1000ms.
We also observe a similar trend in the time to generate first token when
the request rate is beyond 53 req/sec.
Analyzing the LLMRunner artifacts, we also find that medium class shows slightly 
elevated latency variance near saturation points ($\lambda = 68$ and $84$\,req/s), 
where queuing non-determinism amplifies small timing perturbations equally 
for both allocators.

\begin{figure}[t]
\centering
\includegraphics[width=0.81\columnwidth]{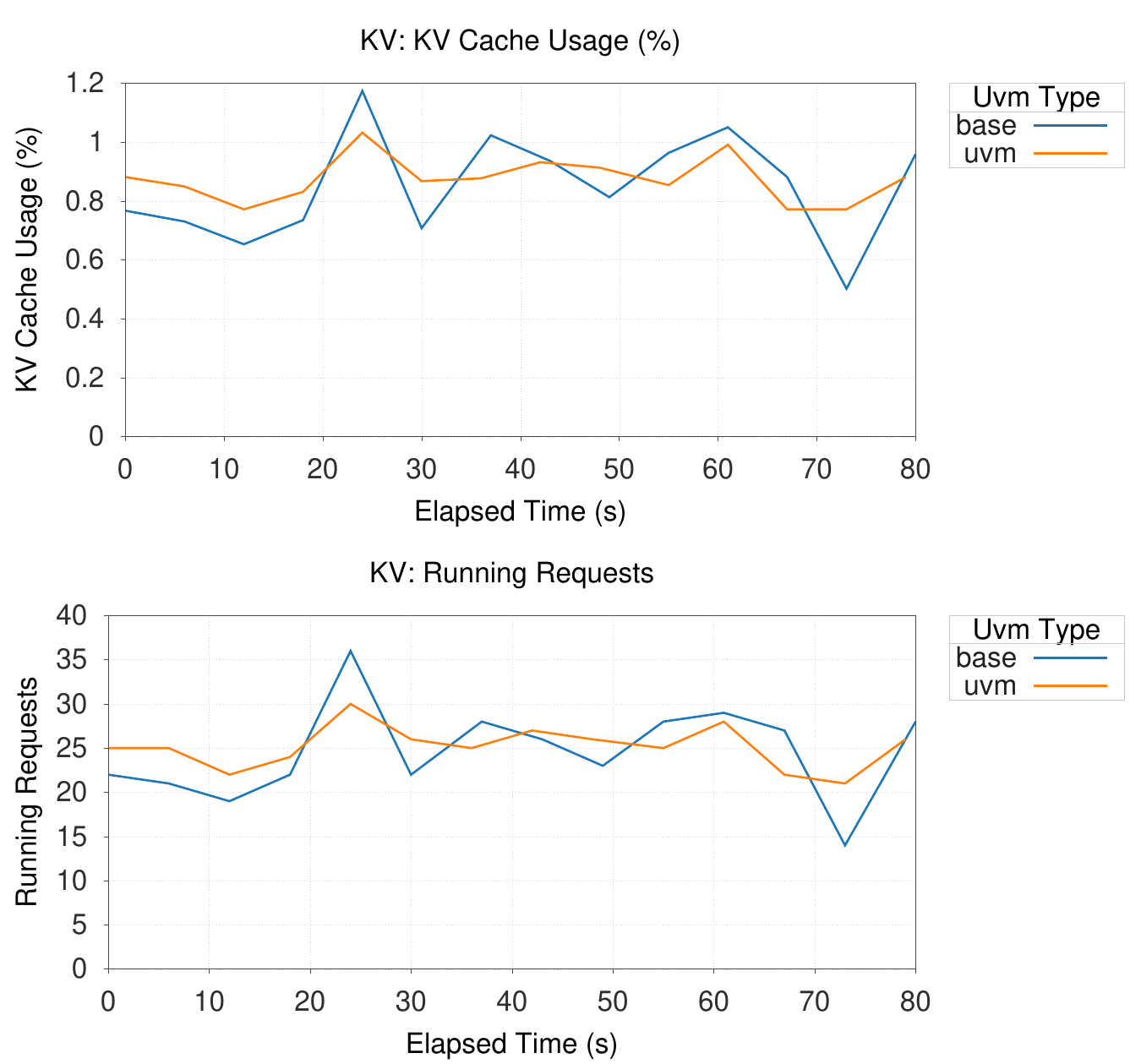}
\caption{Rate mode resource monitor: small class at 75\,req/s ($u{=}0.9$).
  GPU utilization and KV-cache occupancy over elapsed time.  Base and UVM
  resource trajectories are indistinguishable.}
\label{fig:rate-monitor}
\end{figure}

Figure~\ref{fig:rate-monitor} shows GPU utilization (number of active requests in the queue) 
and KV-cache occupancy across elapsed time at one representative rate (75 req/sec).
For both allocators, the GPU resource usage trajectories are similar 
with both allocator modes.

\subsection{Sweeping the Threads Mode}
\label{sec:eval-threads}

\begin{table}[tp]
\caption{Threads mode configuration. Token lengths are drawn from
  normal distributions parameterized as $\mathcal{N}(\mu, \sigma^2)$;
  four worker counts are evaluated per prompt class.}
\label{tab:config-threads}
\centering
\setlength{\tabcolsep}{4pt}
\begin{tabular}{|l|c|c|}
\hline
 & \textbf{Small} & \textbf{Large} \\
\hline
Input tokens  & $\mathcal{N}(128,\,32^2)$ & $\mathcal{N}(512,\,64^2)$ \\
\hline
Output tokens & $\mathcal{N}(64,\,8^2)$   & $\mathcal{N}(256,\,32^2)$ \\
\hline
Workers $N$   & \multicolumn{2}{c|}{32, 48, 64, 80} \\
\hline
\end{tabular}
\end{table}

In threads mode (closed-loop), LLMRunner runs $N$ concurrent worker threads,
each issuing one request, waiting for the reply, and repeating the steps.
To simulates different number of end-users across the experiment
sweep (with zero think time), four concurrency levels are evaluated: $N \in \{32, 48, 64, 80\}$.
Table~\ref{tab:config-threads} lists the token-length distributions and
worker-count sweep used for this experiment.

\begin{figure}[tp]
\centering
\includegraphics[width=0.81\linewidth]{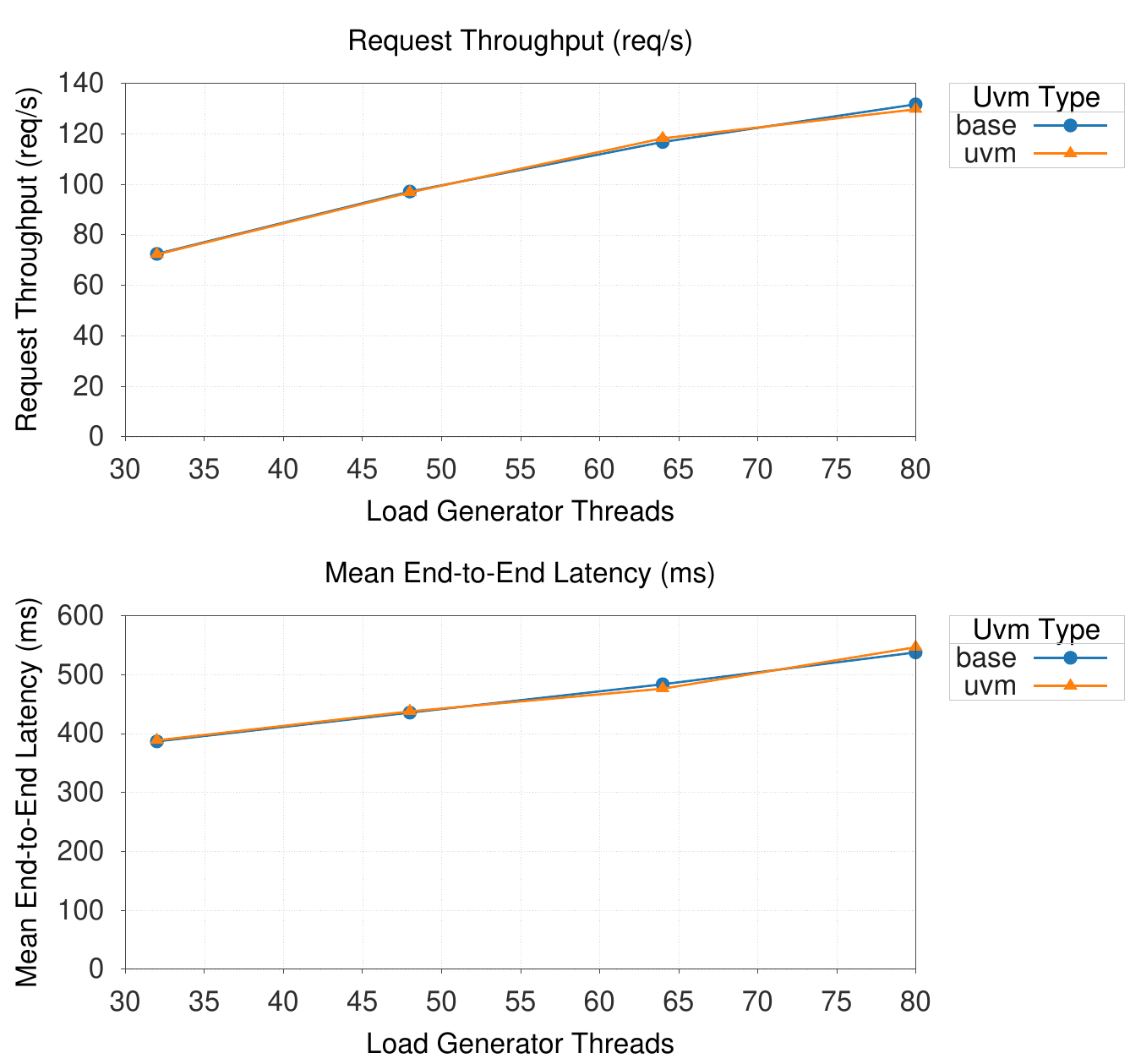}
\caption{Threads mode: small prompt class ($\sim$128\,in / $\sim$64\,out tokens,
  $u{=}0.9$). Throughput and mean end-to-end latency versus worker count $N$.
  Base and UVM traces overlap across all concurrency levels.}
\label{fig:threads-small}
\end{figure}

\begin{figure}[tp]
\centering
\includegraphics[width=0.81\linewidth]{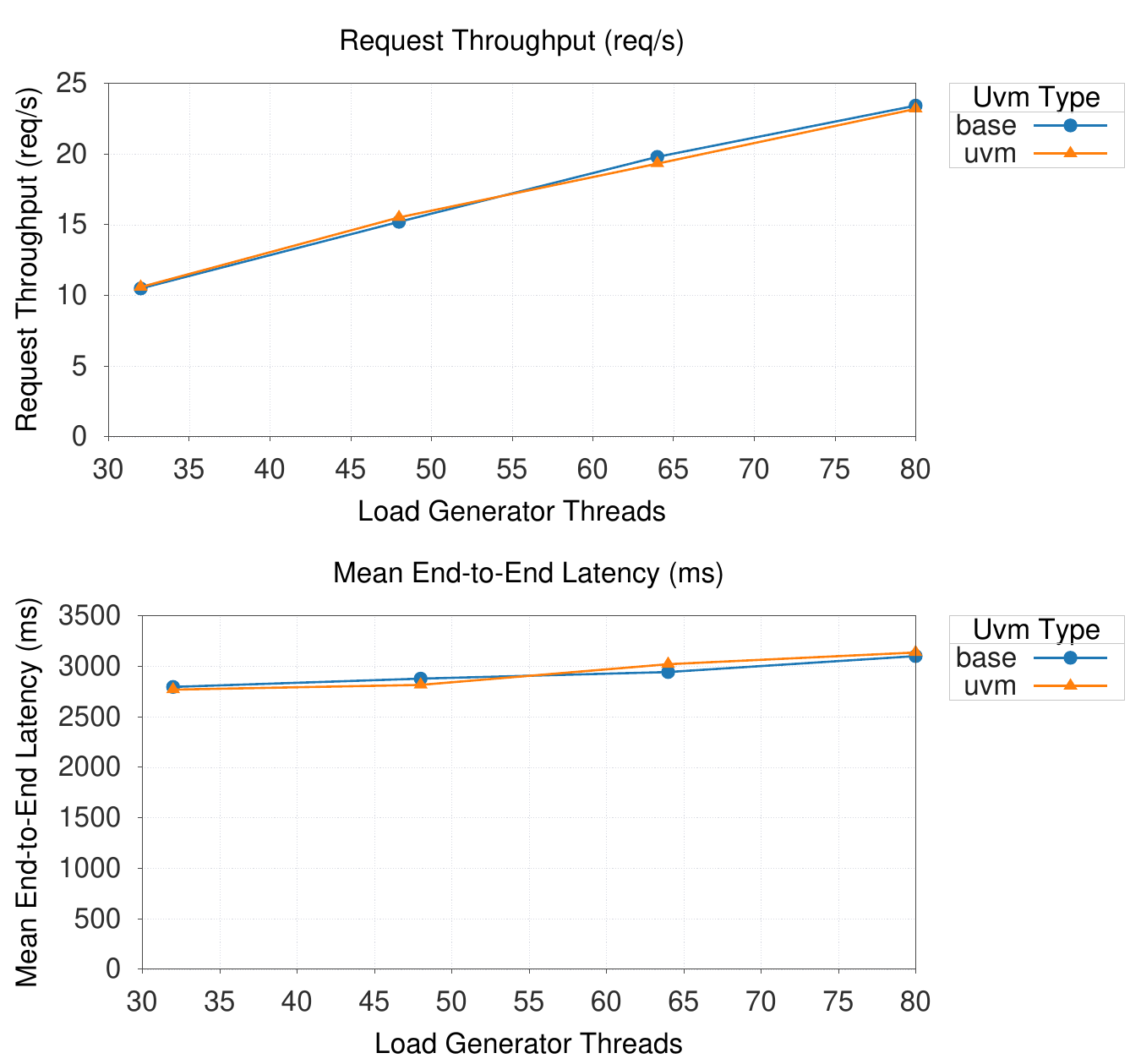}
\caption{Threads mode: medium prompt class ($\sim$512\,in / $\sim$256\,out tokens,
  $u{=}0.9$). Base and UVM throughput and latency remain coincident across
  the concurrency sweep.}
\label{fig:threads-medium}
\end{figure}

Figures~\ref{fig:threads-small} and~\ref{fig:threads-medium} show
achieved throughput and mean end-to-end latency versus worker count for the
small and large classes, respectively.
For the small class, throughput grows from $\sim$72\,req/s at 32 workers to
$\sim$131\,req/s at 80 workers; Base and UVM differ by at most 1.5\,req/s
at any concurrency level, with latency deviations under 2\%.
For the large class, throughput scales from $\sim$10.5 to $\sim$23.4\,req/s
across the worker sweep; UVM and Base diverge by at most 0.3\,req/s in
throughput and $\sim$60\,ms in mean end-to-end latency.
In both classes the closed-loop self-regulation effect is visible---latency
rises gradually with $N$ while throughput continues to grow, consistent with
the theoretical $\lambda_\text{eff} = N / (\bar{T}_\text{resp} +
\bar{T}_\text{think})$ relationship from Section~\ref{sec:load-modes}.

\subsection{Multi-Turn Mode with Real Conversations}
\label{sec:eval-multiturn}

Multi-turn mode in LLMRunner is capable of replaying conversation histories sourced from
real dataset.
For this experiment, we use \textit{allenai/WildChat-1M} dataset~\cite{WildChat},
providing LLM request traces with realistic input lengths and
multi-turn context accumulation.
Two prompt classes exercise different user-pacing behaviors.
Table~\ref{tab:config-multiturn} lists the full configuration;
token-length statistics are derived from the dataset (approximately
128 tokens per turn for both input and output roles on average).
A time budget of 180\,s bounds each run, matching the
\texttt{multi\_turn.duration\_s} plan parameter.
We model two different end-user behaviors---mild and rapid
with different distribution of think time between the turns (Table~\ref{tab:config-multiturn}).

\begin{table}[t]
\caption{Multi-turn mode configuration.
  Think time is sampled per turn from $\mathcal{N}(\mu_t,\,\sigma_t^2)$;
  \textit{Conversations} is the number of distinct sessions loaded
  from the WildChat-1M medium split; token lengths come directly
  from the dataset records.}
\label{tab:config-multiturn}
\centering
\setlength{\tabcolsep}{4pt}
\begin{tabular}{|l|c|c|}
\hline
 & \textbf{Mild} & \textbf{Rapid} \\
\hline
Think time (s) & $\mathcal{N}(5,\;2^2)$ & $\mathcal{N}(1,\;0.1^2)$ \\
\hline
Conversations   & 512  & 2048 \\
\hline
Workers $N$     & \multicolumn{2}{c|}{64,\;128,\;192,\;256} \\
\hline
Dataset         & \multicolumn{2}{c|}{WildChat-1M (medium split)} \\
\hline
Duration        & \multicolumn{2}{c|}{180\,s} \\
\hline
\end{tabular}
\end{table}

\begin{figure}[tp]
\centering
\includegraphics[width=0.81\linewidth]{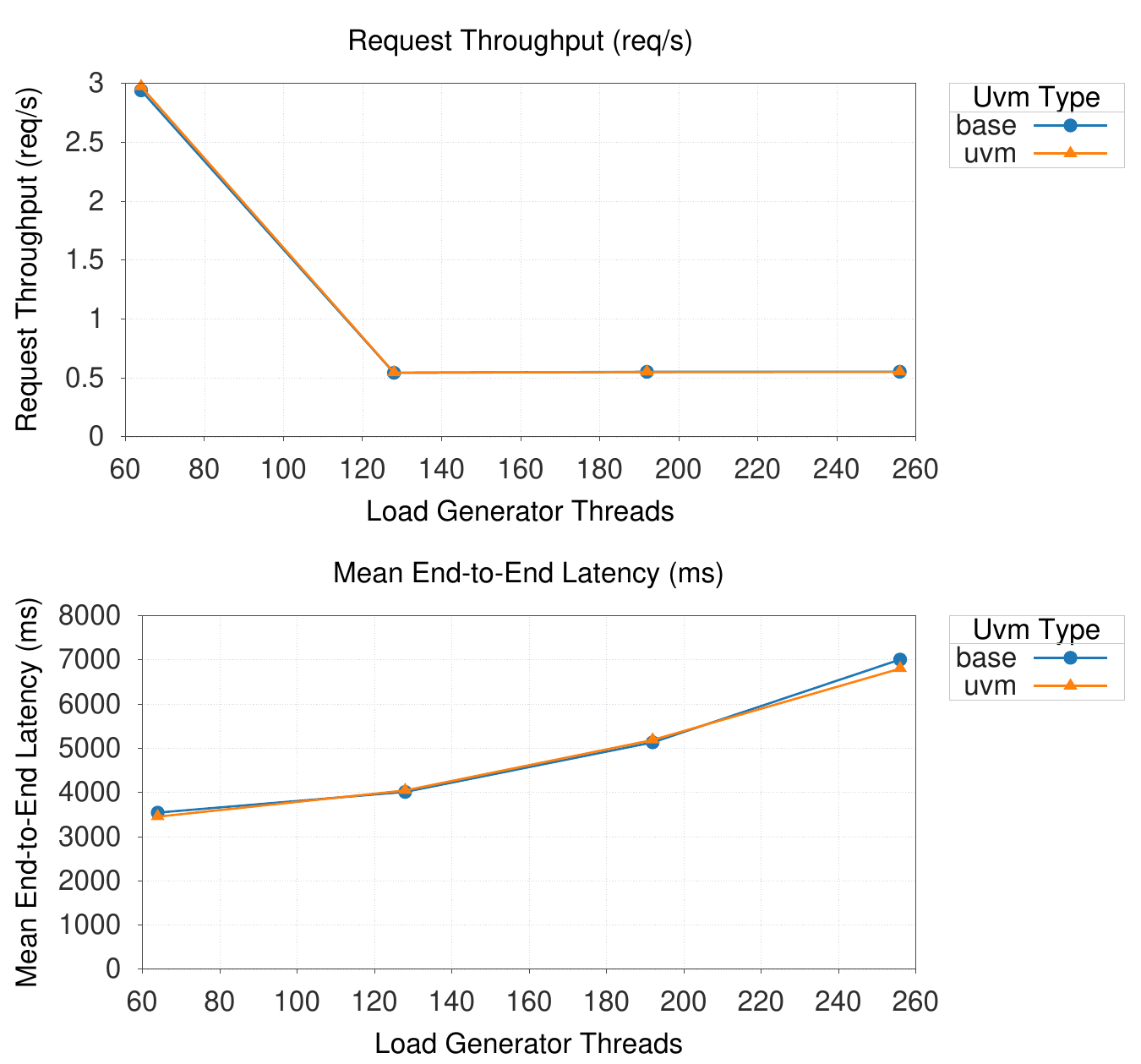}
\caption{Multi-turn mode, mild class
  (think time $\mathcal{N}(5\text{s},\,2^2)$,
  WildChat-1M medium, $u{=}0.9$).
  Turn throughput and mean end-to-end latency versus worker count $N$.
  Base and UVM traces are coincident throughout the sweep.}
\label{fig:wc-mild}
\end{figure}

\begin{figure}[tp]
\centering
\includegraphics[width=0.81\linewidth]{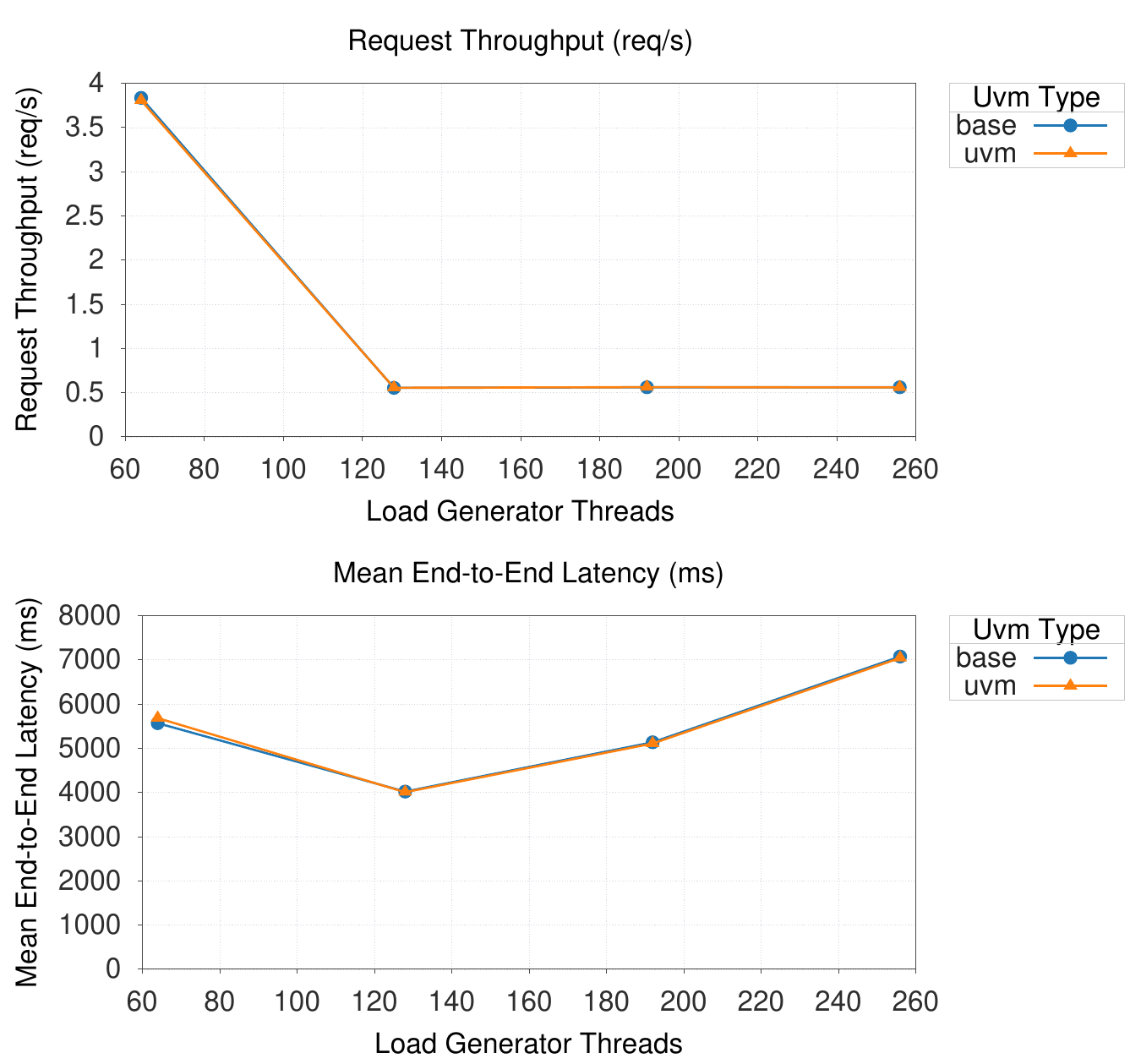}
\caption{Multi-turn mode, rapid class
  (think time $\mathcal{N}(1\text{s},\,0.1^2)$,
  WildChat-1M medium, $u{=}0.9$).
  Base and UVM remain coincident; higher $N$ elevates latency
  without diverging allocator behavior.}
\label{fig:wc-rapid}
\end{figure}

Figures~\ref{fig:wc-mild} and~\ref{fig:wc-rapid} show turn throughput
and mean end-to-end latency versus worker count $N$ for the mild
and rapid classes, respectively.
At $N{=}64$ the mild class achieves $\sim$2.94\,req/s (Base) and
$\sim$2.97\,req/s (UVM), while the rapid class reaches
$\sim$3.83\,req/s and $\sim$3.81\,req/s; mean end-to-end latencies
are $\sim$3.5\,s and $\sim$5.6\,s respectively.
As $N$ increases, concurrent pressure on the KV-cache and scheduler
raises mean latency from $\sim$3.5\,s to $\sim$6.8\,s (mild,
$N{=}256$) and from $\sim$5.6\,s to $\sim$7.0\,s (rapid, $N{=}256$),
consistent with the closed-loop saturation behavior of
Section~\ref{sec:load-modes}.
The rapid class achieves slightly higher per-turn throughput at
$N{=}64$ owing to its shorter inter-turn pause; at higher
concurrency both classes converge to similar completion rates
as server queuing dominates.
In all configurations Base and UVM deliver indistinguishable
performance: turn-throughput divergence is within 2\% and
mean end-to-end latency differs by at most 60\,ms across the
entire concurrency sweep.

\subsection{KV-Cache Rate Tuning}
\label{sec:eval-tune}

To show the utility of LLMRunner, we use the benchmark features to 
find the request rate $\lambda^*$ at which the KV-cache reaches 
a target utilization $u_{\text{KV}}$.
The tuner runs short probes---polling the vLLM
\texttt{/metrics} endpoint between probes---and alternates between
an exponential ramp phase (doubling $\lambda$ until $u_{\text{KV}}$
exceeds the target) and a bisection refine phase (narrowing the
bracket until $|u_{\text{KV}} - u_{\text{target}}| \le \epsilon$).

The experiment uses the medium prompt class
($\mathcal{N}(512,\,64^2)$ input, $\mathcal{N}(256,\,32^2)$ output
tokens) at $u{=}0.3$ GPU memory utilization (lower than the $u{=}0.9$
used in the sweep experiments, so that a moderate request rate
suffices to reach 70\% KV occupancy), targeting 70\%
KV-cache occupancy within a tolerance of $\pm3\%$.
Each probe runs for 30\,s; the p95 of the per-second KV readings
after a 5\,s settle window is used as the probe measurement to
exclude the fill-up transient.

\begin{figure}[tp]
\centering
\includegraphics[width=0.81\linewidth]{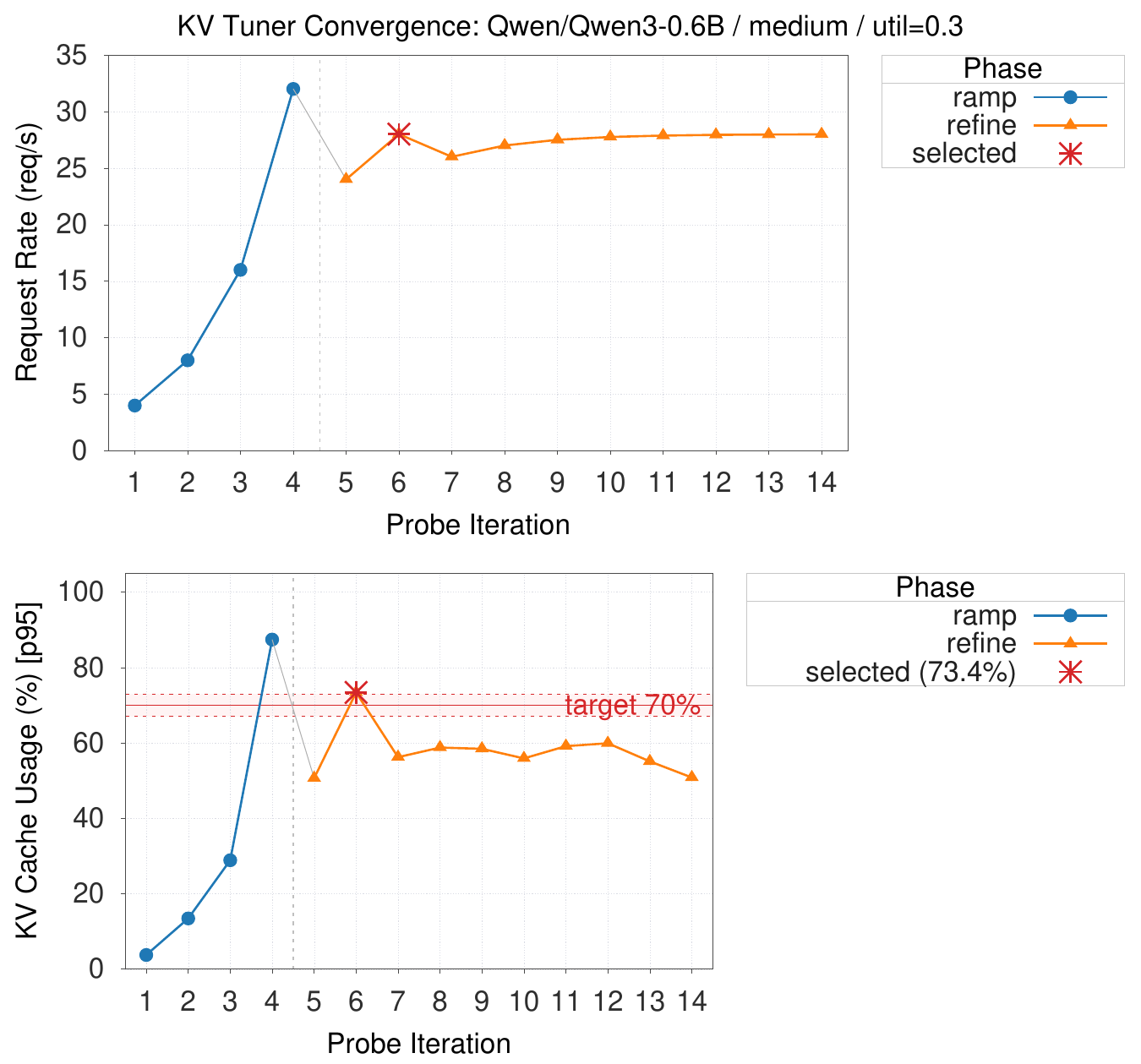}
\caption{KV-cache rate tuner convergence
  (medium class, $u{=}0.3$, target 70\%).
  \textbf{Top:} request rate probed at each iteration; blue
  points are the ramp phase (exponential doubling), orange points
  the bisection refine phase, and the red cross marks the selected
  rate.
  \textbf{Bottom:} measured KV cache usage (p95 across the 30\,s
  probe window); the solid red line is the 70\% target, the dashed
  lines and shaded band mark the $\pm3\%$ tolerance.}
\label{fig:tune-conv}
\end{figure}

Figure~\ref{fig:tune-conv} shows the rate (top panel) and
measured KV utilization (bottom panel) at each probe iteration.
During the ramp phase (probes 1--4) the rate doubles from 4 to
32\,req/s; at 32\,req/s the p95 KV occupancy reaches 87.4\%,
overshooting the 70\% target.
The algorithm switches to bisection, exploring the interval
$[24,\,32]$\,req/s over ten refine probes and converging to
$\lambda^*{=}28.0$\,req/s, which achieves a measured KV of
73.4\%---within 3.4 percentage points of the 70\% target and
inside the declared tolerance band.
The tuner terminates after 14 probes ($\approx$7\,min of probe
time), providing a calibrated operating point that can be
substituted directly into any LLMRunner sweep plan.

%% file: conclusion.tex
\section{Conclusion}
\label{sec:conclusion}
LLMRunner performance testing tool for LLM serving 
generates realistic load consisting of multi-turn conversations 
with configurable end-user parameters such as number of users in the workload loop, 
think-time distributions etc.
We demonstrated the comprehensive set of features---flexible configuration 
to design experiment plans ranging from simple rate-mode experiment 
to closed loop experiments parameterized with different models and 
workload generation methods.
We comparatively studied the performance of LLM serving when KV-Cache
is managed with traditional memory allocation vis-a-vis UVM allocation techniques
using different experiment configuration and reporting features of LLMRunner.
The efficacy of LLMRunner as a performance benchmark tool is demonstrated
through experiments with real conversation multi-turn workloads.

%% file: reference.bib
@inproceedings{PagedAttention,
  author    = {Kwon, Woosuk and Li, Zhuohan and Zhuang, Siyuan and Sheng, Ying
               and Zheng, Lianmin and Yu, Cody Hao and Gonzalez, Joseph and
               Zhang, Hao and Stoica, Ion},
  title     = {Efficient Memory Management for Large Language Model Serving with
               {PagedAttention}},
  booktitle = {Proceedings of the 29th Symposium on Operating Systems Principles},
  pages     = {611--626},
  year      = {2023},
  series    = {SOSP '23},
  publisher = {Association for Computing Machinery}
}

@inproceedings{Orca,
  title     = {Orca: A distributed serving system for {Transformer-Based} generative
               models},
  author    = {Yu, Gyeong-In and Jeong, Joo Seong and Kim, Geon-Woo and Kim,
               Soojeong and Chun, Byung-Gon},
  booktitle = {16th USENIX Symposium on Operating Systems Design and Implementation
               (OSDI 22)},
  pages     = {521--538},
  year      = {2022}
}

@inproceedings{MLPerf,
  title     = {{MLPerf} inference benchmark},
  author    = {Reddi, Vijay Janapa and Cheng, Christine and Kanter, David and
               Mattson, Peter and Schmuelling, Guenther and Wu, Carole-Jean and
               Anderson, Brian and Breughe, Maximilien and Charlebois, Mark and
               Chou, William and others},
  booktitle = {2020 ACM/IEEE 47th Annual International Symposium on Computer
               Architecture (ISCA)},
  pages     = {446--459},
  year      = {2020},
  organization = {IEEE}
}

@misc{LLMPerf,
  title        = {{LLMPerf}: A Framework for Evaluating {LLM} {API}s},
  author       = {Griggs, Cade and Jain, Sid and Khuller, Ankit},
  howpublished = {\url{https://github.com/ray-project/llmperf}},
  year         = {2023}
}

@misc{GenAIPerf,
  title        = {{GenAI-Perf}: Performance Benchmarking Tool for Generative {AI}},
  author       = {{NVIDIA Corporation}},
  howpublished = {\url{https://github.com/triton-inference-server/perf_analyzer/tree/main/genai-perf}},
  year         = {2024}
}

@article{WildChat,
  title   = {{WildChat}: 1M {ChatGPT} interaction logs in the wild},
  author  = {Zhao, Wenting and Ren, Xiang and Hessel, Jack and Cardie, Claire
             and Choi, Yejin and Deng, Yuntian},
  journal = {arXiv preprint arXiv:2405.01470},
  year    = {2024}
}

@article{LLMServingSurvey,
  title   = {Efficient large language model serving: A survey},
  author  = {Miao, Xupeng and Oliaro, Gabriele and Zhang, Zhihao and Cheng, Xinhao
             and Jin, Hongyi and Chen, Tianqi and Jia, Zhihao},
  journal = {arXiv preprint arXiv:2312.15234},
  year    = {2023}
}

@article{llms1,
author = {Naveed, Humza and Khan, Asad Ullah and Qiu, Shi and Saqib, Muhammad and Anwar, Saeed and Usman, Muhammad and Akhtar, Naveed and Barnes, Nick and Mian, Ajmal},
title = {A Comprehensive Overview of Large Language Models},
year = {2025},
publisher = {Association for Computing Machinery},
address = {New York, NY, USA},
volume = {16},
number = {5},
issn = {2157-6904},
url = {https://doi.org/10.1145/3744746},
doi = {10.1145/3744746},
journal = {ACM Transaction on  Intelligent Systems and Technology},
articleno = {106},
numpages = {72},
}

@misc{llms2,
      title={A Comprehensive Survey on Pretrained Foundation Models: A History from BERT to ChatGPT},
      author={Ce Zhou and Qian Li and Chen Li and Jun Yu and Yixin Liu and Guangjing Wang and Kai Zhang and Cheng Ji and Qiben Yan and Lifang He and Hao Peng and Jianxin Li and Jia Wu and Ziwei Liu and Pengtao Xie and Caiming Xiong and Jian Pei and Philip S. Yu and Lichao Sun},
      year={2023},
      eprint={2302.09419},
      archivePrefix={arXiv},
      primaryClass={cs.AI},
      url={https://arxiv.org/abs/2302.09419},
}

@misc{llms3,
      title={A Survey of Large Language Models},
      author={Wayne Xin Zhao and Kun Zhou and Junyi Li and Tianyi Tang and Xiaolei Wang and Yupeng Hou and Yingqian Min and Beichen Zhang and Junjie Zhang and Zican Dong and Yifan Du and Chen Yang and Yushuo Chen and Zhipeng Chen and Jinhao Jiang and Ruiyang Ren and Yifan Li and Xinyu Tang and Zikang Liu and Peiyu Liu and Jian-Yun Nie and Ji-Rong Wen},
      year={2026},
      eprint={2303.18223},
      archivePrefix={arXiv},
      primaryClass={cs.CL},
      url={https://arxiv.org/abs/2303.18223},
}

@inproceedings{closedperf1,
author = {Shams, Mahnaz and Krishnamurthy, Diwakar and Far, Behrouz},
title = {A model-based approach for testing the performance of web applications},
year = {2006},
isbn = {1595935843},
publisher = {Association for Computing Machinery},
address = {New York, NY, USA},
url = {https://doi.org/10.1145/1188895.1188909},
doi = {10.1145/1188895.1188909},
booktitle = {Proceedings of the 3rd International Workshop on Software Quality Assurance},
pages = {54-61},
numpages = {8},
location = {Portland, Oregon},
series = {SOQUA '06}
}

@inproceedings{closedperf2,
author = {Avritzer, Alberto and Kondek, Joe and Liu, Danielle and Weyuker, Elaine J.},
title = {Software performance testing based on workload characterization},
year = {2002},
isbn = {1581135637},
publisher = {Association for Computing Machinery},
address = {New York, NY, USA},
url = {https://doi.org/10.1145/584369.584373},
doi = {10.1145/584369.584373},
booktitle = {Proceedings of the 3rd International Workshop on Software and Performance},
pages = {17-24},
numpages = {8},
location = {Rome, Italy},
series = {WOSP '02}
}

@misc{jmeter,
	title={Apache JMeter},
	Author={Apache},
	url={https://jmeter.apache.org/},
}

@misc{lrp,
	title={OpenText Professional Performance Engineering (LoadRunner Professional)},
	Author={OpenText},
	url={https://www.opentext.com/products/professional-performance-engineering},
}

@ARTICLE{openrist,
  author={George, Shilpa and Eiszler, Thomas and Iyengar, Roger and Turki, Haithem and Feng, Ziqiang and Wang, Junjue and Pillai, Padmanabhan and Satyanarayanan, Mahadev},
  journal={IEEE Pervasive Computing},
  title={OpenRTiST: End-to-End Benchmarking for Edge Computing},
  year={2020},
  volume={19},
  number={4},
  pages={10-18},
  doi={10.1109/MPRV.2020.3028781}
}

@article{faasperf,
author = {Wen, Jinfeng and Chen, Zhenpeng and Zhao, Jianshu and Sarro, Federica and Ping, Haodi and Zhang, Ying and Wang, Shangguang and Liu, Xuanzhe},
title = {SCOPE: Performance Testing for Serverless Computing},
year = {2025},
issue_date = {November 2025},
publisher = {Association for Computing Machinery},
address = {New York, NY, USA},
volume = {34},
number = {8},
issn = {1049-331X},
url = {https://doi.org/10.1145/3717609},
doi = {10.1145/3717609},
journal = {ACM Transaction on Software Engineering Methodology},
month = oct,
articleno = {227},
numpages = {30}
}

@inproceedings{cloudperf1,
author = {He, Sen and Manns, Glenna and Saunders, John and Wang, Wei and Pollock, Lori and Soffa, Mary Lou},
title = {A statistics-based performance testing methodology for cloud applications},
year = {2019},
isbn = {9781450355728},
publisher = {Association for Computing Machinery},
address = {New York, NY, USA},
url = {https://doi.org/10.1145/3338906.3338912},
doi = {10.1145/3338906.3338912},
booktitle = {Proceedings of the 2019 27th ACM Joint Meeting on European Software Engineering Conference and Symposium on the Foundations of Software Engineering},
pages = {188–199},
numpages = {12},
location = {Tallinn, Estonia},
series = {ESEC/FSE 2019}
}

@INPROCEEDINGS{cloudperf2,
  author={He, Sen and Liu, Tianyi and Lama, Palden and Lee, Jaewoo and Kim, In Kee and Wang, Wei},
  booktitle={IEEE/ACM International Conference on Automated Software Engineering (ASE)},
  title={Performance Testing for Cloud Computing with Dependent Data Bootstrapping},
  year={2021},
  pages={666-678},
  doi={10.1109/ASE51524.2021.9678687}
}

@inproceedings{kvc1,
author = {Liu, Yuhan and Li, Hanchen and Cheng, Yihua and Ray, Siddhant and Huang, Yuyang and Zhang, Qizheng and Du, Kuntai and Yao, Jiayi and Lu, Shan and Ananthanarayanan, Ganesh and Maire, Michael and Hoffmann, Henry and Holtzman, Ari and Jiang, Junchen},
title = {CacheGen: KV Cache Compression and Streaming for Fast Large Language Model Serving},
year = {2024},
isbn = {9798400706141},
publisher = {Association for Computing Machinery},
address = {New York, NY, USA},
url = {https://doi.org/10.1145/3651890.3672274},
doi = {10.1145/3651890.3672274},
booktitle = {Proceedings of the ACM SIGCOMM 2024 Conference},
pages = {38-56},
numpages = {19},
location = {Sydney, NSW, Australia},
series = {ACM SIGCOMM '24}
}

@inproceedings{kvc2,
author = {Lee, Wonbeom and Lee, Jungi and Seo, Junghwan and Sim, Jaewoong},
title = {InfiniGen: efficient generative inference of large language models with dynamic KV cache management},
year = {2024},
isbn = {978-1-939133-40-3},
publisher = {USENIX Association},
address = {USA},
booktitle = {Proceedings of the 18th USENIX Conference on Operating Systems Design and Implementation},
articleno = {9},
numpages = {18},
location = {Santa Clara, CA, USA},
series = {OSDI'24}
}

@article{little,
author = {Little, John D. C.},
title = {OR FORUM---Little's Law as Viewed on Its 50th Anniversary},
year = {2011},
issue_date = {May 2011},
publisher = {INFORMS},
address = {Linthicum, MD, USA},
volume = {59},
number = {3},
issn = {0030-364X},
url = {https://doi.org/10.1287/opre.1110.0940},
doi = {10.1287/opre.1110.0940},
journal = {Operation Research},
month = may,
pages = {536-549},
numpages = {14},
}

@inproceedings{uvm1,
author = {Lin, Mao and Feng, Yuan and Cox, Guilherme and Jeon, Hyeran},
title = {Forest: Access-aware GPU UVM Management},
year = {2025},
isbn = {9798400712616},
publisher = {Association for Computing Machinery},
address = {New York, NY, USA},
url = {https://doi.org/10.1145/3695053.3731047},
doi = {10.1145/3695053.3731047},
booktitle = {Proceedings of the 52nd Annual International Symposium on Computer Architecture},
pages = {137-152},
numpages = {16},
series = {ISCA '25}
}

@INPROCEEDINGS {uvm2,
author = { Shin, Hyunkyun and Bang, Seongtae and Park, Hyungwon and Kim, Daehoon },
booktitle = { 2026 IEEE International Symposium on High Performance Computer Architecture (HPCA) },
title = {{ ARIADNE: Adaptive UVM Management for Efficient GPU Memory Oversubscription }},
year = {2026},
pages = {1-15},
doi = {10.1109/HPCA68181.2026.11408564},
publisher = {IEEE Computer Society},
address = {Los Alamitos, CA, USA},
month =Feb
}

@misc{uvmnvidia,
	title={Unified Virtual Memory (UVM)},
	Author={NVIDIA},
	url={https://docs.nvidia.com/ai-enterprise/release-8/latest/infra-software/vgpu/features/uvm.html},
}

@inproceedings{kvco1,
author = {Sun, Chengyu and Xia, Yaqi and Wang, Hulin and Yang, Donglin and Zhou, Xiaobo and Cheng, Dazhao},
title = {JanusQuant: Accurate and Efficient 2-bit KV Cache Quantization for Long-Context Inference},
year = {2026},
isbn = {9798400723100},
publisher = {Association for Computing Machinery},
address = {New York, NY, USA},
url = {https://doi.org/10.1145/3774934.3786428},
doi = {10.1145/3774934.3786428},
booktitle = {Proceedings of the 31st ACM SIGPLAN Annual Symposium on Principles and Practice of Parallel Programming},
pages = {301–314},
numpages = {14},
location = {Sydney, NSW, Australia},
series = {PPoPP '26}
}

@inproceedings{kvco2,
author = {Hong, Yang and Yang, Junlong and Peng, Bo and Yao, Jianguo},
title = {REPA: Reconfigurable PIM for the Joint Acceleration of KV Cache Offloading and Processing},
year = {2026},
isbn = {9798400723599},
publisher = {Association for Computing Machinery},
address = {New York, NY, USA},
url = {https://doi.org/10.1145/3779212.3790212},
doi = {10.1145/3779212.3790212},
booktitle = {Proceedings of the 31st ACM International Conference on Architectural Support for Programming Languages and Operating Systems, Volume 2},
pages = {1622–1639},
numpages = {18},
location = {USA},
series = {ASPLOS '26}
}
